\documentclass[11pt]{article}
\usepackage{eurosym}
\usepackage{amsfonts}
\usepackage{amssymb}
\usepackage{amsmath}
\usepackage{latexsym}
\usepackage{float,subfigure,color,subfigure}
\usepackage{graphicx}
\usepackage{verbatim}

\usepackage{amsfonts}
\usepackage{amssymb}
\usepackage{amsmath}
\usepackage{latexsym}

\newcommand{\Sfrac}[2]{{\textstyle{\frac{#1}{#2}}}}

\newcommand{\ohf}{\frac{1}{2}}

\newcommand{\scrO}{{\mathcal{O}}}

\newcommand{\R}{\mathbb{R}}

\newcommand{\tx}{\tilde{x}}

\newcommand{\ttt}{\tilde{t}}
\newcommand{\teta}{\tilde{\eta}}

\newcommand{\tD}{\tilde{D}}
\newcommand{\tH}{\tilde{H}}
\newcommand{\tu}{\tilde{u}}
\newcommand{\tr}{\tilde{r}}
\newcommand{\ts}{\tilde{s}}
\newcommand{\tK}{\tilde{K}}

\newcommand{\ta}{\alpha}

\newcommand{\sca}{-\kappa/\mu^\nu}

\DeclareMathOperator{\sech}{sech}

\author{
Daulet Moldabayev\thanks{\texttt{daulet.moldabayev@math.uib.no}}, \,
Henrik Kalisch\thanks{\texttt{henrik.kalisch@math.uib.no}} \\ 
{\small Department of Mathematics, University of Bergen} \\
{\small Postbox 7800, 5020 Bergen, Norway} \\
\\
Denys Dutykh\thanks{\texttt{Denys.Dutykh@univ-savoie.fr}} \\
{\small LAMA, UMR5127, CNRS - Universit\'{e} de Savoie, Campus Scientifique} \\
{\small 73376 Le Bourget-du-Lac Cedex, France} \\  }

\title{The Whitham Equation as a Model for Surface Water Waves}

\begin{document}

\maketitle

\begin{abstract}
The Whitham equation was proposed as an alternate model equation for
the simplified description of uni-directional wave motion at the
surface of an inviscid fluid. As the Whitham equation incorporates the full linear dispersion relation
of the water wave problem, it is thought to provide a more faithful
description of shorter waves of small amplitude than traditional long wave models such as the KdV equation.

In this work, we identify a scaling regime in which the Whitham equation can be derived from the Hamiltonian
theory of surface water waves. 
The Whitham equation is integrated numerically, and it is shown that the equation 
gives a close approximation of inviscid free surface dynamics as described by the Euler equations.
The performance of the Whitham equation as a model for free surface dynamics is 
also compared to two standard free surface models: the KdV and the BBM equation.
It is found that in a wide parameter range of amplitudes and wavelengths,
the Whitham equation performs on par with or better than both the KdV and BBM equations.
\end{abstract}

\section{Introduction}
In its simplest form, the water-wave problem concerns the flow of an incompressible inviscid
fluid with a free surface over a horizontal impenetrable bed. In this situation, the fluid flow 
is described by the Euler equations with appropriate boundary conditions,
and the dynamics of the free surface are of particular interest in the solution of this problem.

There are a number of model equations which
allow the approximate description of the evolution of the free surface
without having to provide a complete solution of the fluid flow
below the surface.
In the present contribution, interest is focused on the derivation and evaluation of
a nonlocal water-wave model known as the Whitham equation.
The equation is written as
\begin{equation}\label{dimWhitham}
\eta_t+\frac{3}{2}\frac{c_0}{h_0}\eta\eta_x+K_{h_0} \! \ast \eta_{x}=0,
\end{equation}
where the convolution kernel $K_{h_0}$ is given in terms of the Fourier transform by
\begin{equation}\label{symbol}
\mathcal{F} K_{h_0} (\xi) = {\textstyle{ \sqrt{\frac{g\tanh (h_0\xi)}{\xi}}}}, 
\end{equation}
$g$ is the gravitational acceleration, $h_0$ is the undisturbed depth of the fluid, 
and $c_0 = \sqrt{g h_0}$ is the corresponding long-wave speed. 
The convolution can be thought of as a Fourier multiplier operator, and \eqref{symbol}
represents the Fourier symbol of the operator. 

The Whitham equation was proposed by Whitham \cite{Wh1} as an alternative to the
well known Korteweg-de Vries (KdV) equation
\begin{equation}\label{KdV}
 \eta_t + c_0 \eta_x + \frac{3}{2}\frac{c_0}{h_0}\eta\eta_x+ \frac{1}{6} c_0 h_0^2 \eta_{xxx} =0.
\end{equation}
The validity of the KdV equation as a model for  surface water waves
can be described as follows.
Suppose a wave field with a prominent amplitude $a$ and characteristic wavelength
$l$ is to be studied. 
The KdV equation is known to produce a good approximation of the evolution of the waves 
if the amplitude of the waves is small 
and the wavelength is large when compared to the undisturbed depth,
and if in addition, the two non-dimensional quantities $a / h_0$ and 
$h_0^2 / l^2$ are of similar size.
The latter requirement can be written in terms of the Stokes number as
$$
\mathcal{S} = \frac{a l^2}{h_0^3} \sim 1.
$$ 
While the KdV equation is a good model for surface waves
if $\mathcal{S} \sim 1$,
one notorious problem with the KdV equation is that it does not model
accurately the dynamics of shorter waves.
Recognizing this shortcoming of the KdV equation, Whitham proposed to use
the same nonlinearity as the KdV equation, but couple it with a linear term
which mimics the linear dispersion relation of the full water-wave problem.
Thus, at least in theory, 
the Whitham equation can be expected to yield a description of the dynamics 
of shorter waves which is closer to the solutions of the more fundamental
Euler equations which govern the flow.

The Whitham equation has been studied from a number of vantage points 
during recent years. In particular, the existence of traveling and solitary waves
has been established in \cite{EGW,EK1}. Well posedness of a similar equation
was investigated in \cite{LannesBOOK,LS}, and a model with variable depth has
been studied numerically in \cite{AMP}.
Moreover, it has been shown
in \cite{HJ,Sanford}
that periodic solutions of the Whitham equation feature modulational instability 
for short enough waves in a similar way as small-amplitude periodic wave
solutions of the water-wave problem. However, it appears that the 
performance of the Whitham equation in the description of surface water
waves has not been investigated so far.

The purpose of the present article is to 
give an asymptotic derivation of the Whitham equation
as a model for surface water waves, and to
confirm Whitham's expectation that the equation is a fair model
for the description of time-dependent surface water waves.
For the purpose of the derivation, 
we introduce an exponential scaling regime in which the
Whitham equation can be derived asymptotically from an
approximate Hamiltonian principle for surface water waves.
In order to motivate the use of this scaling, note that the KdV equation
has the property that wide classes of initial data 
decompose into a number of solitary waves 
and small-amplitude dispersive residue \cite{AS}.
For the KdV equations,
solitary-wave solutions are known in closed form, and are given by
\begin{equation} \label{kdvsol}
\eta= \frac{a}{h_0}~ \mbox{sech}^2\left( {\textstyle \sqrt{ \frac{3a}{4h_0^3}}(x-ct)} \right)
\end{equation}
for a certain wave celerity $c$.
These waves clearly comply with the amplitude-wavelength relation 
$a / h_0 \sim h_0^2 / l^2 $
which was mentioned above.
It appears that the Whitham equation - as indeed do many other nonlinear dispersive equations - 
also has the property that broad classes of initial data rapidly 
decompose into ordered trains of solitary waves (see Figure 1). 
Quantifying the amplitude-wavelength relation for these solitary waves
yields an asymptotic regime which is expected to be relevant to the validity of the
Whitham equation as a water wave model.

As the curve fit in the right panel of Figure 1 shows, the relationship 
between wavelength and amplitude of the Whitham solitary waves 
can be approximately described by the relation
$\frac{a}{h_0} \sim e^{-\kappa(l/h_0)^\nu}$
for certain values of $\kappa$ and $\nu$.
Since the Whitham solitary waves are not known in exact form,
the values of $\kappa$ and $\nu$ have to be found numerically.
Then one may define a Whitham scaling regime 
\begin{equation}\label{Whithamnumber}
\mathcal{W}(\kappa, \nu) = \frac{a}{h_0}e^{\kappa(l/h_0)^\nu } \sim 1,
\end{equation}
and it will be shown in sections 2 and 3 that this scaling
can be used advantageously in the derivation of the Whitham equation.
The derivation proceeds by examining the Hamiltonian
formulation of the water-wave problem due to Zhakarov, Craig and Sulem \cite{CS,Zh},
and by restricting to wave motion which is predominantly in the direction
of increasing values of $x$.
The approach is similar to the method of 
\cite{CG}, but relies on the new relation \eqref{Whithamnumber}.
\begin{figure}[]
\centering
\includegraphics[width=0.48\textwidth]{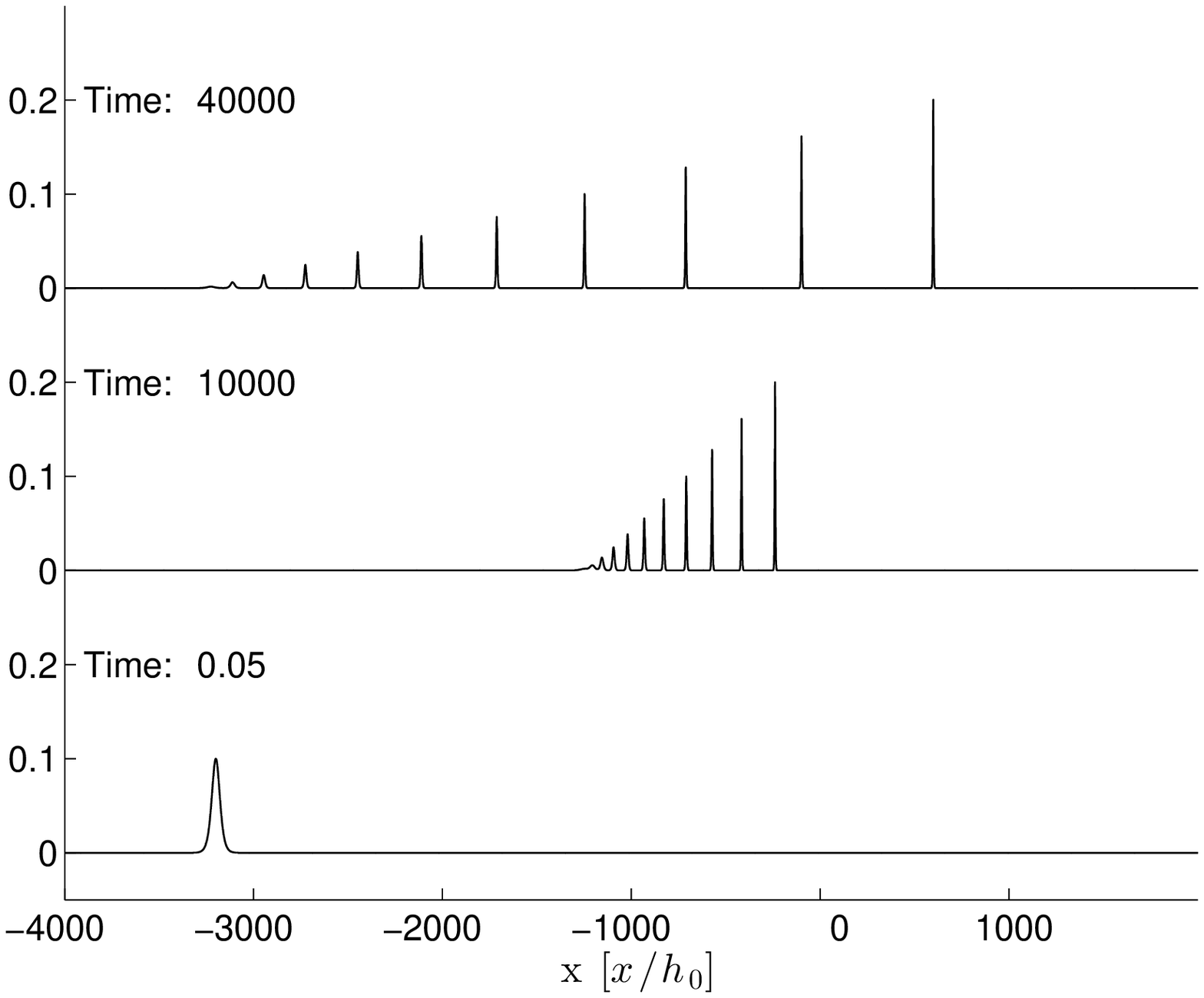}
\includegraphics[width=0.48\textwidth]{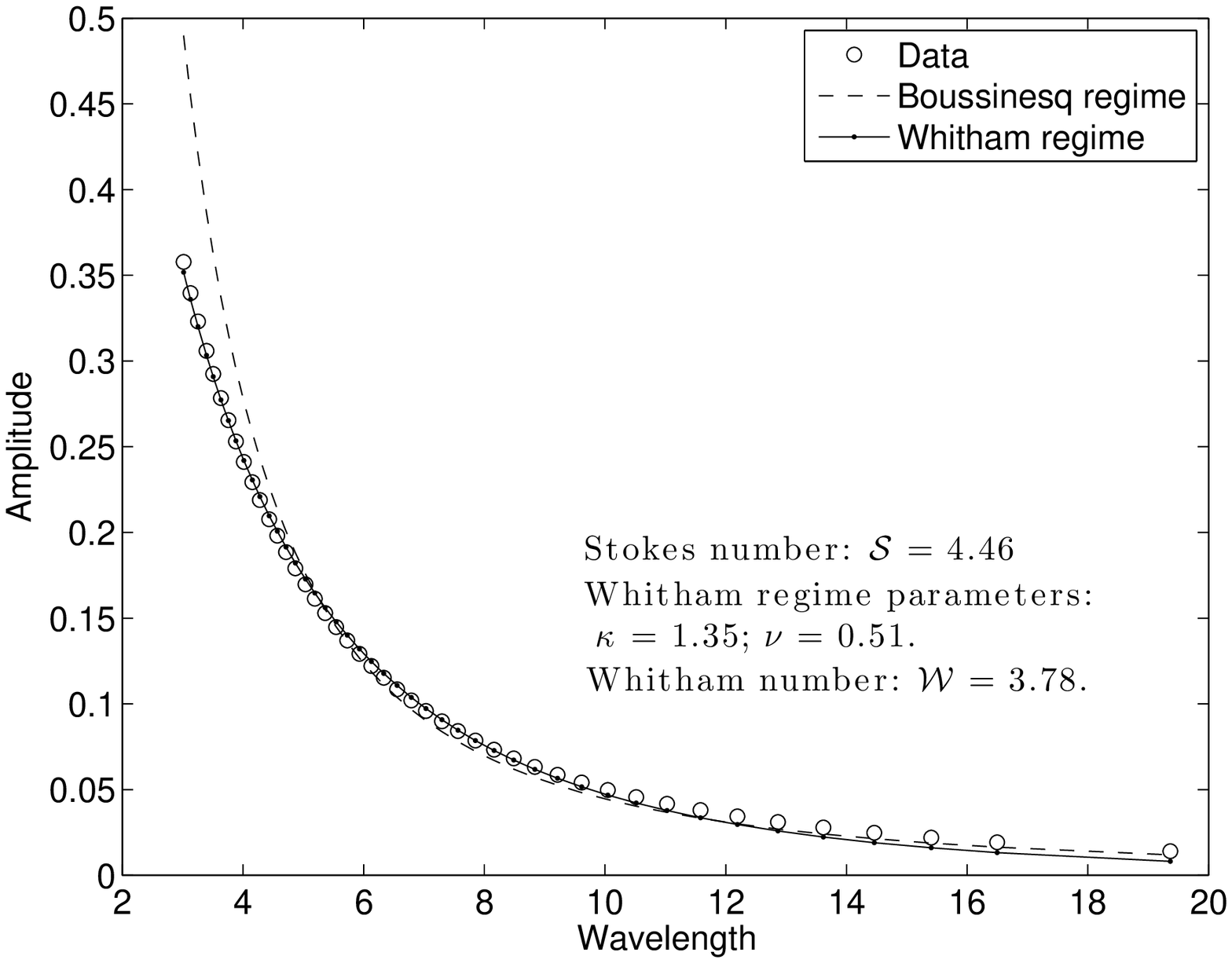}
\caption{Left panel: Formation of solitary waves of the Whitham equation from Gaussian initial data. 
Right panel: Curve fit for the Whitham regime and for the Boussinesq regime
to amplitude/wavelength data from Whitham solitary waves. The wavelength is defined
as $l = \frac{1}{a} \int_{-\infty}^{\infty}\eta(x)dx$. 
}
\label{fitting-figure}
\end{figure} 

First, in Section 2, a Whitham system is derived which allows for two-way propagation
of waves. The Whitham equation is found in Section 3.
Finally, in Section 4, a comparison of modeling properties
of the KdV and Whitham equations is given. 
The comparison also includes
the regularized long-wave equation
\begin{equation}\label{BBM}
\eta_t + c_0 \, \eta_x + \frac{3}{2}\frac{c_0}{h_0} \, \eta \, \eta_x 
                    - \frac{1}{6} h_0^2 \, \eta_{xxt} = 0,
\end{equation}
which was put forward in \cite{Pe} and studied in depth in \cite{BBM}, 
and which is also known as the BBM or PBBM equation. 
The linearized dispersion relation 
of this equation is not an exact match to the dispersion relation of the full water-wave problem,
but it is much closer than the KdV equation, and it might also be expected that 
this equation may be able to model shorter waves more
successfully than the KdV equation. However, as will be seen, 
solutions of the Whitham
equation appear to give a closer approximation to solutions of the full Euler
equations than either \eqref{KdV} or \eqref{BBM} in most cases investigated.

\section{Derivation of evolution systems of Whitham type}
%
%
The surface water-wave problem 
is generally described by the Euler equations with slip conditions at
the bottom, and kinematic and dynamic boundary conditions at
the free surface. Assuming weak transverse effects, the unknowns
are the surface elevation $\eta(x,t)$, the horizontal and vertical 
fluid velocities $u_1(x,z,t)$ and $u_2(x,z,t)$, respectively, and the pressure $P(x,z,t)$.
If the assumption of irrotational flow is made, then
a velocity potential $\phi(x,z,t)$ can be used.
In order to nondimensionalize the problem, the undisturbed depth $h_0$ 
is taken as a unit of distance, 
and the parameter $\sqrt{h_0/g}$ as a unit of time.
For the remainder of this article, all variables appearing in the
water-wave problem are considered as being non-dimensional.
The problem is posed on a domain
$\left\{(x,z)^T \in \R^2 | -1 < z < \eta(x,t) \right\}$
which extends to infinity in the positive and negative $x$-direction.
Due to the incompressibility of the fluid, the potential then satisfies 
Laplace's equation in this domain.
The fact that the fluid cannot penetrate the bottom is expressed by
a homogeneous Neumann boundary condition at the flat bottom.
Thus we have 
\begin{eqnarray*}
\phi_{xx} + \phi_{zz} = 0 & \mbox{in} &  -1 < z < \eta(x,t) \\
\phi_z = 0              & \mbox{on} &  z=-1.
\end{eqnarray*}
The pressure is eliminated with help
of the Bernoulli equation, and the free-surface boundary conditions are formulated
in terms of the potential and the surface excursion by
\[
\left.
\begin{array}{rc}
\eta_t+\phi_x\eta_x-\phi_z & =0, \\
\phi_t+\frac{1}{2} \big( \phi^2_x+\phi^2_z \big) + \eta & = 0, 
\end{array}
\right\}
\mbox{on} \ z=\eta(x,t).
\]
The total energy of the system is given by the sum of kinetic energy
and potential energy, and normalized such that the potential energy
is zero when no wave motion is present at the surface.
Accordingly the Hamiltonian function for this problem is
\begin{equation}
H = \int_\R \int_0^\eta z \, dz dx + \int_\R \int_{-1}^\eta \ohf |\nabla \phi|^2 \, dz dx.
\end{equation}
Defining the trace of the potential at the free surface as
$\Phi(x,t) = \phi(x,\eta(x,t),t)$,
one may integrate in $z$ in the first integral and use the divergence theorem 
on the second integral in order to arrive at the formulation
\begin{equation}\label{Hamiltonian-eta-phi}
H  = \ohf \int_\R \left[ \eta^2 + \Phi G(\eta) \Phi \right] \, dx.
\end{equation}
This is the Hamiltonian formulation of the water wave problem as found in
\cite{CS,Petrov,Zh}, and written in terms
of the Dirichlet-Neumann operator $G(\eta)$.
As shown in \cite{Nicholls}, the Dirichlet-Neumann operator
can be expanded as a power series as
\begin{equation}
G(\eta)\Phi = \sum_{j=0}^\infty G_j(\eta) \Phi.
\end{equation}
In order to proceed, we need to understand the first few terms in this series.
As shown in \cite{CG,CS}, the first two terms in this series
can be written with the help of the operator $D = - i \partial_x$ as 
\begin{equation*}
G_0(\eta)= D\tanh(D),
\qquad 
G_1(\eta)= D\eta D - D\tanh(D) \eta D\tanh(D). 
\end{equation*}
Note that it can be shown that the terms $G_j(\eta)$ for $j\ge 2$ are of quadratic or
higher-order in $\eta$, and will therefore not be needed in the current
analysis.

It will be convenient for the present purpose to formulate the Hamiltonian
in terms of the dependent variable $u = \Phi_x$.
To this end, we define the operator $\mathcal{K}(\eta)$ by
\begin{equation*}
G(\eta) = D \mathcal{K}(\eta) D.
\end{equation*}
As was the case with $G(\eta)$, the operator $\mathcal{K}(\eta)$ 
can be expanded in a Taylor series around zero as
\begin{equation}
\mathcal{K}(\eta) \xi = \sum_{j=0}^\infty \mathcal{K}_j(\eta) \xi~, 
\quad \quad \mathcal{K}_j(\eta) = D^{-1}G_j(\eta)D^{-1}.
\end{equation}
In particular, note that $\mathcal{K}_0 = \frac{\tanh{D}}{D}$.
In non-dimensional variables, we write the operator with the integral
kernel $K_{h_0}$ as $K = \sqrt\frac{\tanh{D}}{D}$, so that we have
$\mathcal{K}_0 = K^2$.
The Hamiltonian is then expressed as 
\begin{equation}\label{Hamiltonian-eta-u}
H  = \ohf \int_\R \left[ \eta^2 + u \mathcal{K}(\eta) u \right] \, dx.
\end{equation}
The water-wave problem can then be written as a Hamiltonian system using
the variational derivatives of $H$ and posing the Hamiltonian equations
\begin{equation}\label{Hamilton-system-eta-phi}
\eta_t = - \partial_x \frac{\delta H}{\delta u}, 
\quad \quad 
u_t = - \partial_x \frac{\delta H}{\delta \eta}.
\end{equation}
This system is not in canonical form as the associated structure map $J_{\eta,u}$ is symmetric:
\begin{equation*}
J_{\eta,u} =
\begin{pmatrix}
0 & -\partial_x \\
-\partial_x & 0
\end{pmatrix}.
\end{equation*}
We now proceed to derive a system of equations which is similar to the Whitham equation
\eqref{dimWhitham}, but allows bi-directional wave propagation. This system will be
a stepping stone on the way to a derivation of \eqref{dimWhitham}, but may also be
of independent interest.
Consider a wavefield having a characteristic wavelength $l$ and a characteristic amplitude $a$. 
Taking into account the nondimensionalization, the two scalar parameters $\lambda = l / h_0$ 
and $\alpha = a / h_0$ appear.
In order to introduce the long-wave and small amplitude approximation into the non-dimensional problem,
we use the scaling $\tx = \frac{1}{\lambda} x$, and $\eta = \alpha \teta$. 
This induces the transformation 
$\tD = \lambda D = - \lambda i \partial x.$
If the energy is nondimensionalized in accord with the nondimensionalization mentioned
earlier, then the natural scaling for the 
Hamiltonian is $\tH = \ta^2 H$.
In addition, the 
unknown $u$ is scaled as $u = \ta \tu$.
The scaled Hamiltonian \eqref{Hamiltonian-eta-u} is then written as
\begin{multline*}
\tH = \ohf \int_{\R} \teta^2 \, dx
 + \ohf \int_{\R} \tu \left[ 1 - \Sfrac{1}{3} \lambda^{-2} \tD^2 + \cdots \right] \tu \, dx  
 + \frac{\ta}{2} \int_{\R}  \teta  \tu^2 \, dx
\\
 - \frac{\ta}{2} \int_{\R}  \tu \left[ \lambda^{-1} \tD - \Sfrac{1}{3} \lambda^{-3}\tD^3 +  \cdots \right] \teta 
   \left[ \lambda^{-1} \tD - \Sfrac{1}{3} \lambda^{-3}\tD^3 + \cdots \right] \tu \, dx.
\end{multline*}
\noindent 
Let us now introduce the small parameter $\mu= \frac{1}{\lambda}$,
and assume for simplicity that $\ta = e^{- \kappa / \mu^\nu}$, which corresponds to the
case where $\mathcal{W}(\kappa,\nu) = 1$. 
Then the Hamiltonian can be written in the following form:
\begin{multline*}
\tH = \ohf \int_{\R} \teta^2 \, dx
        + \ohf \int_{\R} \tu \left[ 1 - \Sfrac{1}{3} \mu^{2} \tD^2 + \cdots \right] \tu \, dx  
+ \frac{e^{-\kappa/\mu^\nu}}{2} \int_{\R}  \teta  \tu^2 \, dx
\\
- \frac{e^{-\kappa/\mu^\nu}}{2} \int_{\R}  \tu \left[ \mu \tD - \Sfrac{1}{3} \mu^{3}\tD^3 +  \dots\right]
                      \teta \left[ \mu \tD - \Sfrac{1}{3} \mu^{3}\tD^3 + \cdots \right] \tu \, dx.
\end{multline*}
\noindent
Disregarding terms of order $\scrO(\mu^2 e^{-\kappa/\mu^\nu})$, but not of order
$\scrO(e^{\sca})$ yields the expansion
\begin{equation}
\tH = \ohf \int_{\R} \teta^2 \, dx 
    + \ohf \int_{\R} \tu \left[ 1 - \Sfrac{1}{3} \mu^{2} \tD^2 + \dots \right] \tu \, dx  
    + \frac{e^{\sca}}{2} \int_{\R}  \teta  \tu^2 \, dx.   
\end{equation}
\noindent
Note that by taking $\mu$ small enough, an arbitrary number of terms of algebraic order 
in $\mu$ may be kept in the asymptotic series, 
so that the truncated version of the Hamiltonian in dimensional 
variables may be written as
\begin{equation}\label{Hamiltonian-eta-u-truncated}
H   = \ohf \int_\R \left[ \eta^2 + u \mathcal{K}_0^N (\eta) u + u \eta u \right] \, dx dz.
\end{equation}
However, the difference between $\mathcal{K}_0$ and $\mathcal{K}_0^N$
is below the order of approximation, so that it is possible
to formally define the truncated Hamiltonian with
$\mathcal{K}_0$ instead of $\mathcal{K}_0^N$.
\noindent 
Hence, the Whitham system is obtained from the 
Hamiltonian \eqref{Hamiltonian-eta-u-truncated} as follows:
\begin{align}
\label{sys1}
\eta_t = - \partial_x \frac{\delta H}{\delta u} 
       &= -  \mathcal{K}_0 u_x - (\eta u)_x, \\
\label{sys2}
u_t = -\partial_x \frac{\delta H}{\delta \eta}
       &= - \eta_x - u u_x.
\end{align}
\noindent 
One may also derive a higher-order equation by keeping terms of order 
$\scrO(\mu^2 e^{\sca})$, but discarding terms of order $\scrO(\mu^4 e^{\sca})$.
In this case we find the system
\begin{align*}
\eta_t & = -  \mathcal{K}_0 u_x - (\eta u)_x - (\eta u_x)_{xx}, \\
u_t    & = - \eta_x - u u_x + u_x u_{xx}.
\end{align*}
%
%
%
\section{Derivation of evolution equations of Whitham type}
%
In order to derive the Whitham equation for uni-directional wave propagation,
it is important to understand how solutions of the Whitham system \eqref{sys1}-\eqref{sys2}
can be restricted to either left or right-going waves.
As it will turn out, if  $\eta$ and $u$ are such that $\eta = K u$, 
then this pair of functions represents a solution
of \eqref{sys1}-\eqref{sys2} which is propagating to the right. 
%
Indeed, let us analyze the relation between $\eta$ and $u$ in the linearized Whitham system
\begin{align}
\label{linsys1}
\eta_t & = -  \mathcal{K}_0 u_x, \\
\label{linsys2}
u_t    & = - \eta_x.
\end{align}
Considering a solution of the system \eqref{linsys1}-\eqref{linsys2} in the form
\begin{equation}
\eta(x,t) = A e^{(i\xi x - i \omega t)}, \qquad u(x,t) = B e^{(i\xi x - i \omega t)}.
\end{equation}
\noindent 
gives rise to the matrix equation
\begin{equation}
\begin{pmatrix}
-\omega & \frac{\tanh{\xi}}{\xi} \xi \\
\xi  & -\omega
\end{pmatrix}
\begin{pmatrix}
A \\
B
\end{pmatrix} =
\begin{pmatrix}
0 \\
0
\end{pmatrix}.
\label{matrix-equation}
\end{equation}
If existence of a nontrivial solution of this system is to be guaranteed,
the determinant of the matrix has to be zero, so that we have
$\omega^2 - \frac{\tanh{\xi}}{\xi} \xi^2 = 0$. Defining the phase speed
as $c = \omega / \xi$, we obtain the dispersion relation
\begin{equation}
c = \pm \sqrt{{\textstyle \frac{\tanh{\xi}}{\xi}}}.
\end{equation}
The choice of $c > 0$ corresponds to right-going wave solutions of the system 
\eqref{linsys1}-\eqref{linsys2},
and the relation between $\eta$ and $u$ can be deduced from \eqref{linsys2}.
%
%
%
%
Accordingly, it is expedient to separate solutions
into a right-going part $r$ and a left-going part $s$
which are defined by
\begin{align*}
r = \ohf (\eta + K  u), \qquad
s = \ohf (\eta - K  u). 
\end{align*}
According to the transformation theory detailed in \cite{CGK}, 
if the unknowns $r$ and $s$ are used instead of $\eta$ and $u$,
the structure map changes to 
\begin{equation}
J_{r,s} = \left( \frac{\partial F}{\partial (\eta, u)} \right) J_{\eta, u}  
\left( \frac{\partial F}{\partial (\eta, u)} \right)^T
= 
\begin{pmatrix}
-\ohf \partial_x K & 0\\
0 & \ohf \partial_x K
\end{pmatrix}.
\end{equation}
\noindent 
We now use the same scaling for both dependent and independent variables as before.
Thus we have $r = \alpha\tr$ and $s = \alpha\ts$. 
%
%
%
%
The Hamiltonian is written in terms of $\tr$ and $\ts$ as
\begin{multline*}
\tH = \ohf \int_{\R} (\tr+\ts)^2 \, dx
\\
 + \ohf \int_{\R} \tK^{-1}(\tr-\ts) \left[ 1 - \Sfrac{1}{3} \mu^2 \tD^2 + \cdots \right] \tK^{-1}(\tr-\ts) \, dx
 + \frac{\alpha}{2} \int_{\R}  (\tr+\ts) \left(\tK^{-1}(\tr-\ts)\right)^2 \, dx
\\
 - \frac{\alpha}{2} \int_{\R}  \tK^{-1}(\tr-\ts) \left[ \mu \tD - \Sfrac{1}{3} \mu^3\tD^3 +  \cdots \right] (\tr+\ts)
   \left[ \mu \tD - \Sfrac{1}{3} \mu^3\tD^3 + \cdots \right] \tK^{-1}(\tr-\ts) \, dx. 
\end{multline*}
%
%
%
%
%
Following the transformation rules, the structure map transforms to
$J_{\tr,\ts} = 1/\ta^2 J_{r,s}$. 
In addition, the time scaling $t = \lambda \ttt$ is employed. 
Since the focus is on right-going solutions, the equation to be considered is
\begin{equation}
\lambda \tr_{\ttt} 
= -\frac{1}{2\ta^2}\lambda\partial_{\tx} \tK \left[ \frac{\delta \big( \ta^2 \tH \big)}{\delta \tr} \right].
\end{equation}
So far, this equation is exact. If we now assume that 
$s$ is of the order of
$\scrO(\mu^2 e^{\sca})$, then the equation for $\tr$ is
\begin{multline*}
\tr_{\ttt} = -\frac{1}{2}\partial_{\tx} \big[ 1 - \Sfrac{1}{6} \mu^2 \tD^2 + \cdots \big] 
           \bigg\{ 2 \tr + \frac{\alpha}{2} \left( \big[1 + \Sfrac{1}{6} \mu^2 \tD^2 + \cdots \big] 
           \tr \right)^2 
\\
+ \alpha \big[ 1 + \Sfrac{1}{6} \mu^2 \tD^2 + \cdots \big] 
   \left( \tr \, \big[1 + \Sfrac{1}{6} \mu^2 \tD^2 + \cdots \big]  \tr  \right)  
- \frac{\alpha}{2} \left( \big[ \mu \tD - \Sfrac{1}{3} \mu^3 \tD^3 + \cdots \big] 
    \big[ 1 + \Sfrac{1}{6} \mu^2 \tD^2 + \cdots \big]  \tr  \right)^2
\\
- \alpha \big[ \mu \tD - \Sfrac{1}{3} \mu^3 \tD^3 + \cdots \big] \big[1 + \Sfrac{1}{6} \mu^2 \tD^2 + \cdots \big] 
 \left( \tr \big[ \mu \tD - \Sfrac{1}{3} \mu^3 \tD^3 + \cdots \big] [1 + \Sfrac{1}{6} \mu^2 \tD^2 
    + \cdots \big] \tr \right) \bigg\} + \scrO(\alpha \mu^2).
\end{multline*}
As in the case of the Whitham system, we use $\ta = \scrO(e^{- \kappa / \mu^\nu})$,
and disregard terms of order $\scrO(\mu^2 e^{\sca})$, but not of order $\scrO(e^{\sca})$.
This procedure yields the Whitham equation \eqref{dimWhitham} 
which is written in nondimensional variables as
\begin{equation*}
r_t = -K r_x - \frac{3}{2} r r_x.
\end{equation*}
As was the case for the system found in the previous section, it is also 
possible here to include terms of order $\scrO(\mu^2 e^{-\kappa/\mu^\nu})$, resulting
in the higher-order equation
\begin{equation*}
r_t = -K r_x - \frac{3}{2} r r_x - \frac{13}{12} r_x r_{xx} - \frac{5}{12} r r_{xxx}.
\end{equation*}
%
%
%
\section{Numerical results}
%
In this section, the performance of the Whitham
equation as a model for surface water waves 
is compared 
to both the KdV equation \eqref{KdV}
and to the BBM equation \eqref{BBM}. 
For this purpose initial data are imposed, 
the Whitham, KdV and BBM equations are solved numerically,
and the solutions are compared to numerical solutions of the full Euler equations
with free-surface boundary conditions. We continue to work in normalized
variables, such as stated in the beginning of Section 2.

The numerical treatment of the three model equations is 
by a standard pseudo-spectral scheme,
such as explained in \cite{EK2,FW} for example. For the time stepping,
an efficient fourth-order implicit method developed in \cite{FruSan} is used. 
The numerical treatment of the free-surface problem for the Euler equations
is based on a conformal mapping of the fluid domain into a rectangle.
In the time-dependent case, this method has roots in the work of Ovsyannikov \cite{Ovsyannikov}, 
and was later used in \cite{DyachenkoZakharov} and \cite{LHCh}. 
The particular method used for the numerical experiments reported here 
is a pseudo-spectral scheme which is detailed in \cite{MDC}.

\begin{figure}
\centering     
\subfigure{\includegraphics[width=0.69\textwidth]{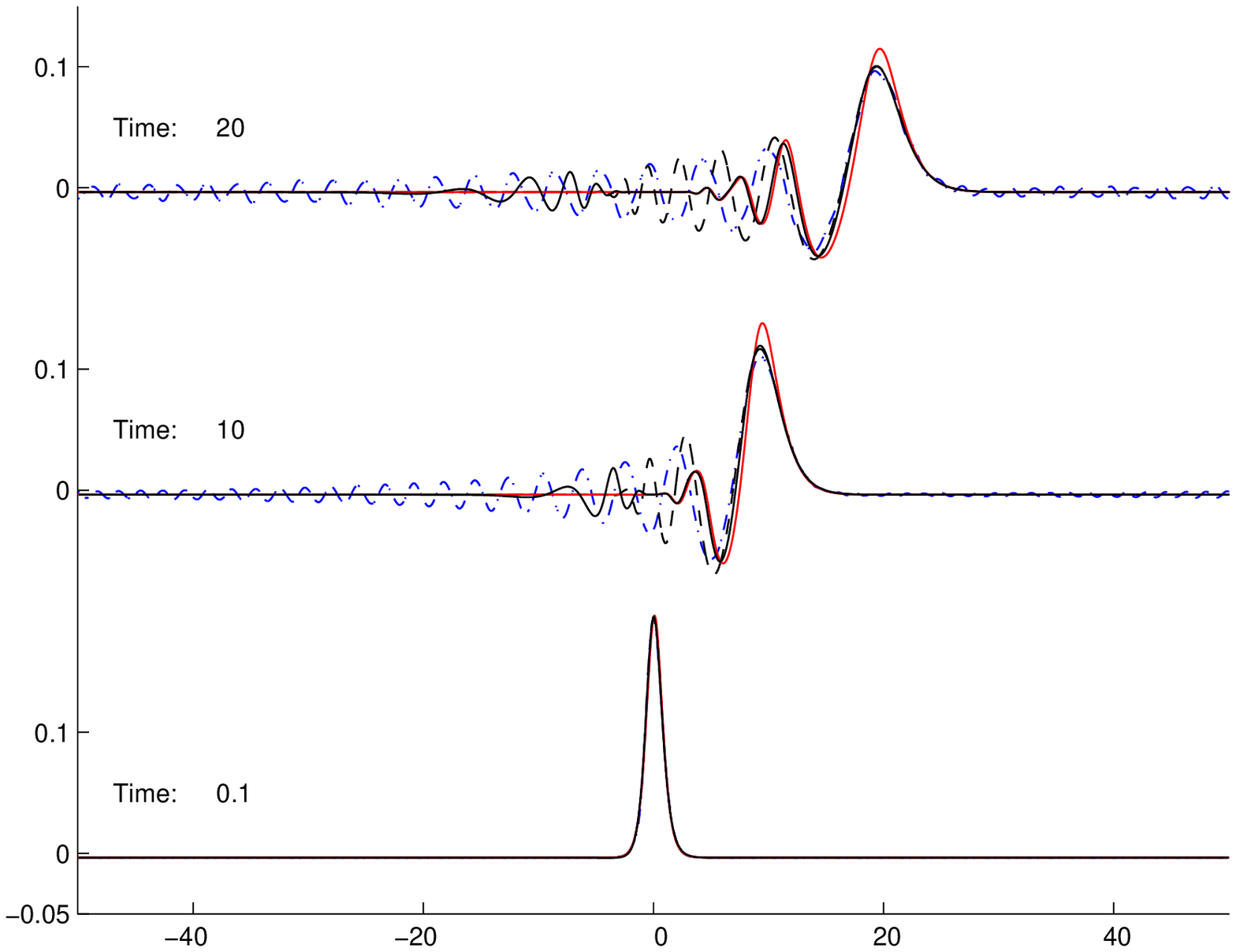}}
\subfigure{\includegraphics[width=0.69\textwidth]{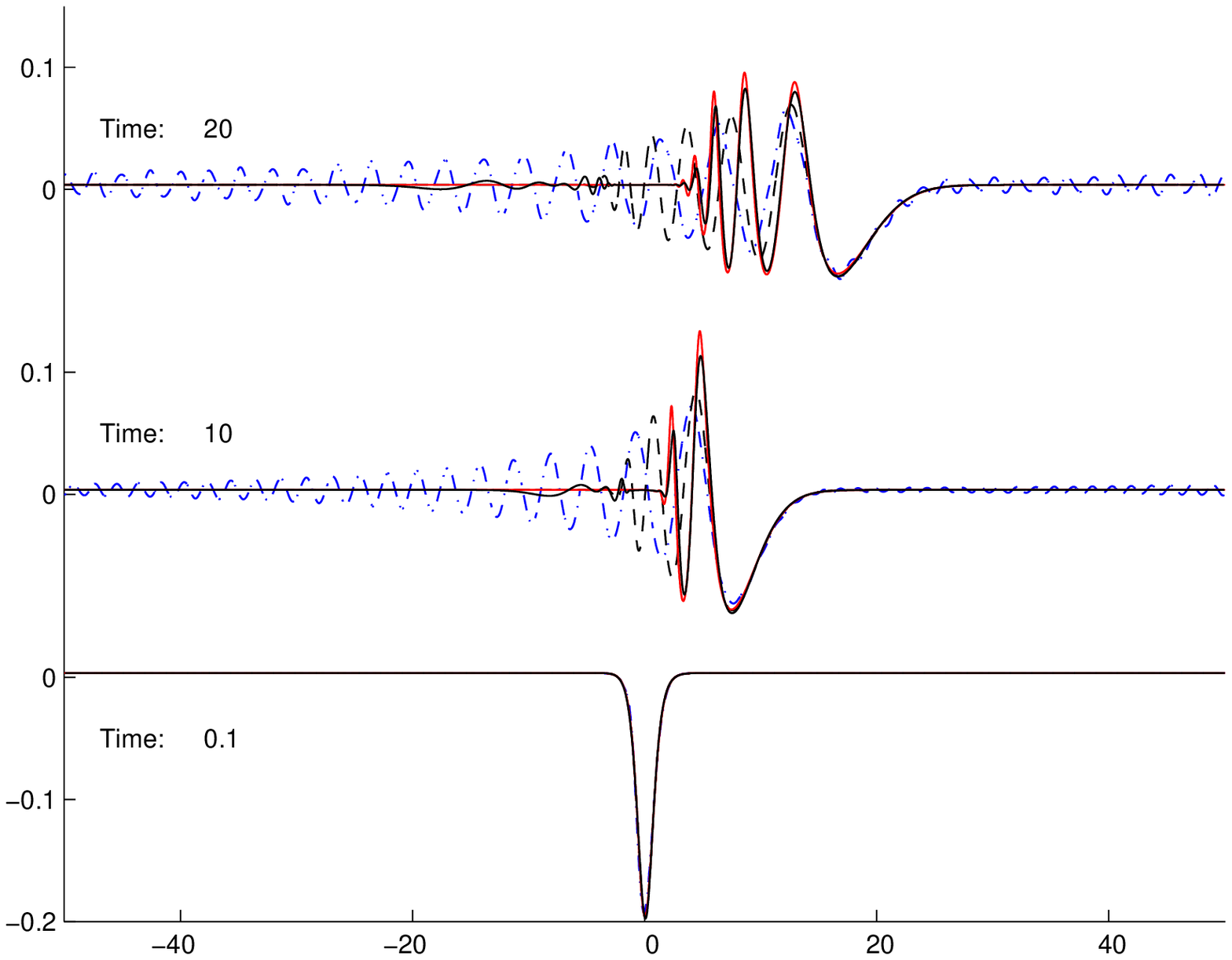}}
\caption{\small Wave profiles at three different times: \textcolor{black}{\textbf{--}} Euler, \textcolor{blue}{\textbf{-$\cdot$-}} KdV, \textbf{- -} BBM, \textcolor{red}{\textbf{--}} Whitham. Experiment B: $\mathcal{S} = 1,~ \alpha = 0.2,~ \lambda = \sqrt{5}$. Upper panel: positive case; lower panel: negative case. 
Horizontal axis: $x/h_0$, vertical axis: $z/h_0$. Snapshots are given at nondimensional time 
$t/ \sqrt{h_0/g}.$}
\label{fig2}
\end{figure}

Initial conditions for the Euler equations are chosen in such a way that 
the solutions are expected to be right moving. This is achieved by
posing an initial surface disturbance $\eta_0(x)$ together with the
trace of the potential $\Phi(x) = \int_0^x \eta_0(x')\, dx'.$
In order to normalize the data, we choose $\eta_0(x)$ in such a way that
the average of $\eta_0(x)$ over the computational domain is zero.
The experiments are performed 
with several different amplitudes $\alpha$ and wavelengths $\lambda$
(for the purpose of this section, we define the 
wavelength $\lambda$ as the distance between 
the two points $x_1$ and $x_2$ at which $\eta_0(x_1) = \eta_0(x_2) = \alpha/2$).
Both positive and negative initial disturbances are considered.
While disturbances with positive main part have been studied widely, 
an initial wave of depression is somewhat more exotic, but nevertheless important,
as shown for instance in \cite{Hammack}.
A summary of the experiments' settings is given in Table 1.
Experiments run with an initial wave of elevation are
labeled as {\it positive},
and experiments run with an initial wave of depression are
labeled as {\it negative}.
The domain for the computations is $-L\leq x \leq L$, with $L = 50$. 
The function initial data in the {\it positive} cases is given by
\begin{equation}
\eta_0(x) =  \alpha \sech^2(f(\lambda)x)-C, \label{test_1}
\end{equation}
where 
\begin{equation*}
f(\lambda) = \frac{2}{\lambda}\log\left({\textstyle \frac{1+\sqrt{1/2}}{\sqrt{1/2}}}\right), 
\ \mbox{ and } \
C = \frac{1}{L}\frac{\alpha}{f(\lambda)}\tanh \left({\textstyle \frac{L}{f(\lambda)}}\right).
\end{equation*}
and $C$ and $f(\lambda)$ are chosen so that 
$\int_{-L}^{L}\eta_0(x)dx = 0$, and
the wavelength $\lambda$ is the distance between the 
two points $x_1$ and $x_2$ at which $\eta_0(x_1) = \eta_0(x_2) = a/2$.
The velocity potential in this case is given by
\begin{equation}
\label{velocity1}
\Phi(x) = \frac{\alpha}{f(\lambda)}\tanh(f(\lambda)x)-Cx. 
\end{equation}
In the {\it negative} case, the initial data are given by
\begin{equation*}
\eta_0(x) = - \alpha \sech^2(f(\lambda)x) + C. 
\end{equation*}
The definitions for $f(\lambda)$ and $C$ are the same, and the velocity potential is
\begin{equation*}
\Phi(x) = -\frac{\alpha}{f(\lambda)}\tanh(f(\lambda)x)+Cx. 
\end{equation*}
\begin{table}
\begin{center}
	\begin{tabular}{| c | c | c | c |}
	\hline
	Experiment & Stokes number & $\alpha$ & $\lambda$ \\
	\hline
	A	& 0.2  &  	 0.1 &	$\sqrt{2}$ 	\\	
	B	& 0.2  & 	 0.2 &	1 			\\
	C	& 1 	   & 	 0.1 &	$\sqrt{10}$	\\
	D	& 1    & 	 0.2 &	$\sqrt{5}$	\\	
	E	& 5    & 	 0.1 &	$\sqrt{50}$ 	\\
	F	& 5    & 	 0.2 &	5			\\
	\hline
	\end{tabular}
\end{center}
\caption{Summary of the Stokes number, nondimensional amplitude and
nondimensional wavelength of the initial data used in the numerical
experiments.}
\end{table}
In Figure \ref{fig2}, the time evolution of a wave with an initial narrow peak
and one with an initial narrow depression at the center is shown. 
The amplitude
is $\alpha=0.2$, and the wavelength is $\lambda = \sqrt{5}$. The time evolution
according to the Euler, Whitham, KdV and BBM equations are shown. It appears
that the KdV equation produces a significant number of spurious oscillations,
the BBM equation also produces a fair number of spurious oscillations, and
the Whitham equation produces fewer small oscillations than Euler equations.
Moreover, while the highest peak in the upper panel in Figure \ref{fig2} is
underpredicted by the KdV and BBM equation, the Whitham equation
produces a peak which is slightly too high.
In the case of an initial depression, the Whitham equation also
produces some peaks which are too high, but on the other hand,
both the KdV and BBM equations introduce a phase error in the
main oscillations.

\begin{figure}
\centering
\subfigure{\includegraphics[width=0.39\textwidth]{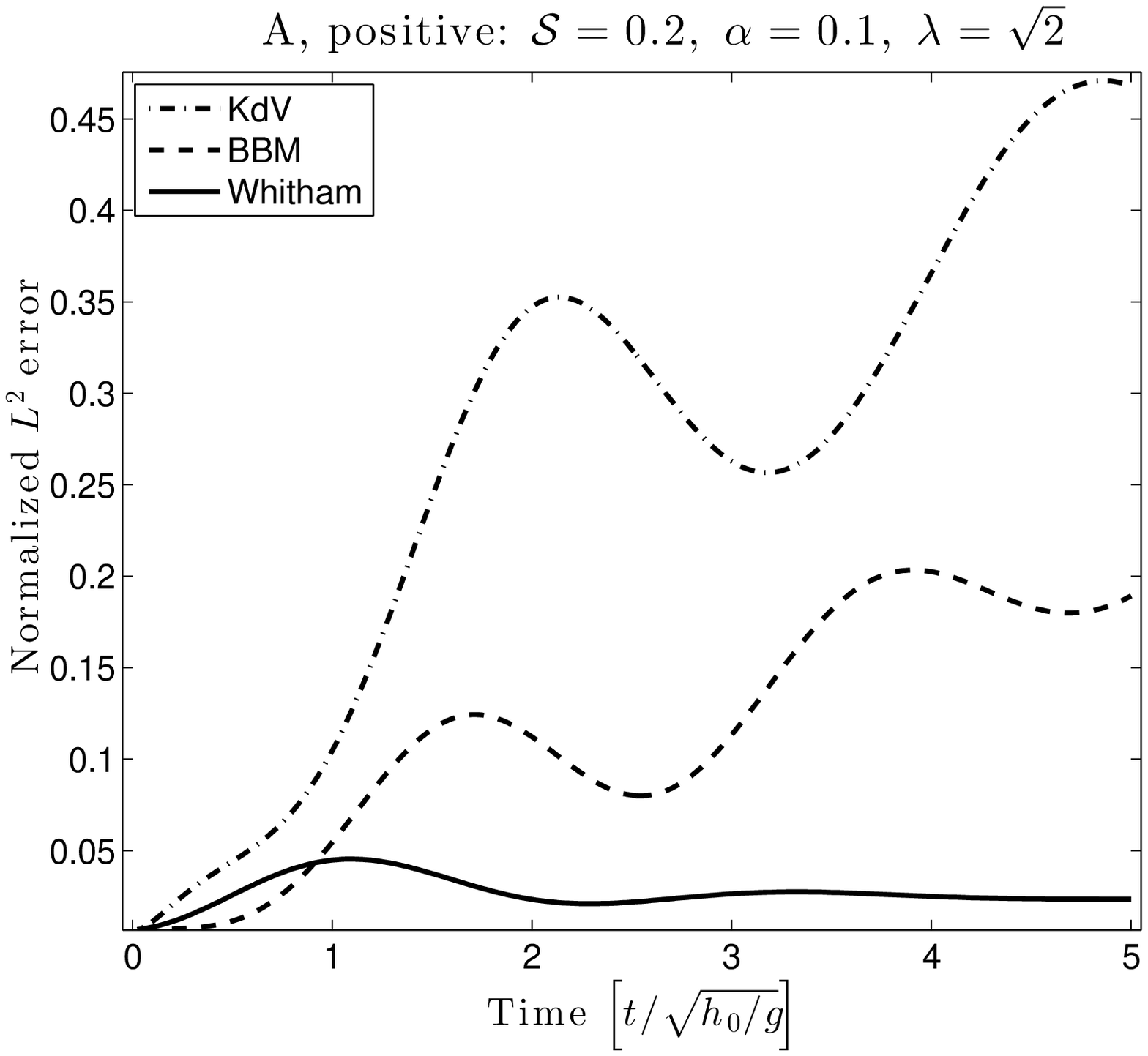}}~~~~
\subfigure{\includegraphics[width=0.39\textwidth]{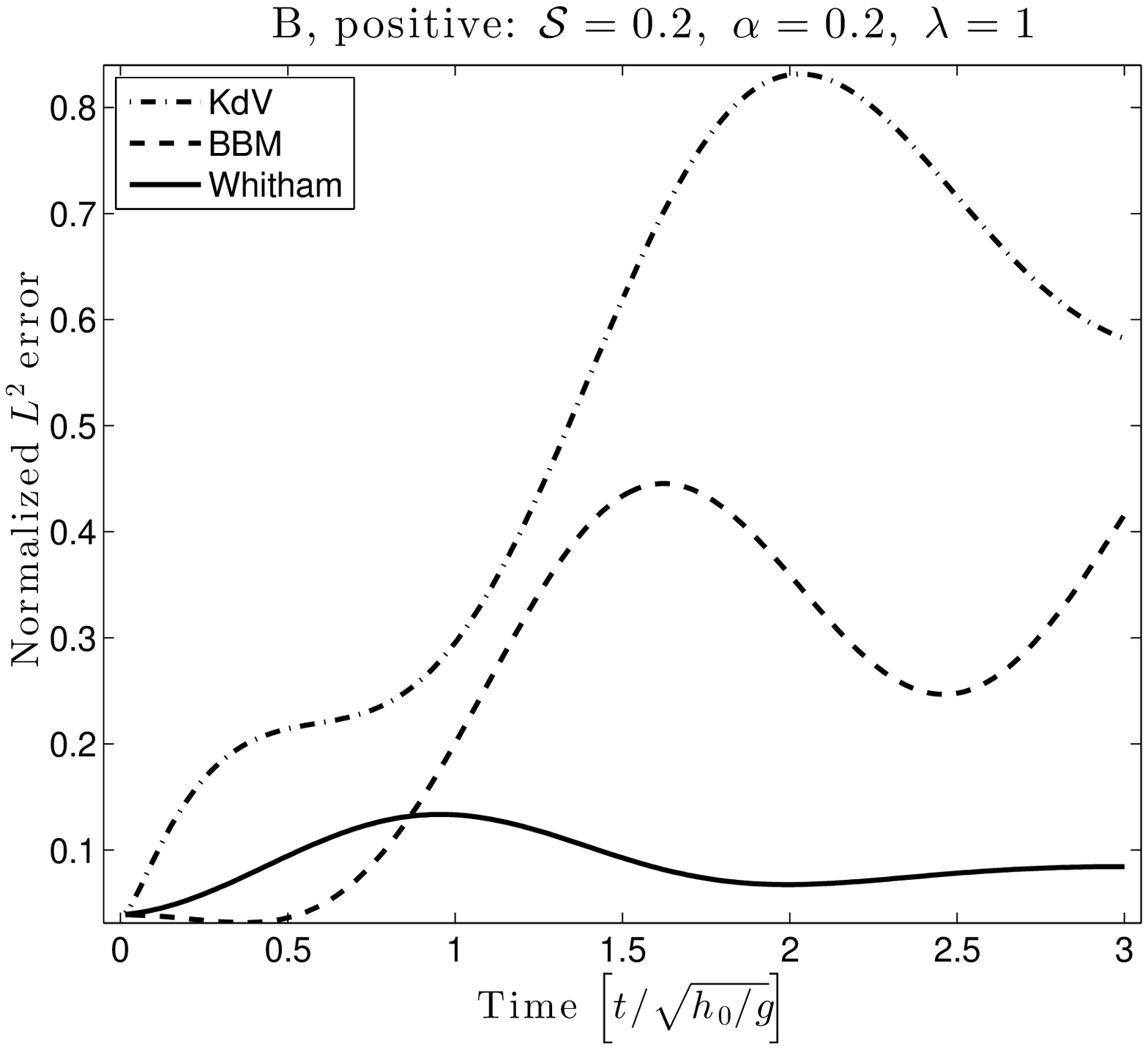}}
\subfigure{\includegraphics[width=0.39\textwidth]{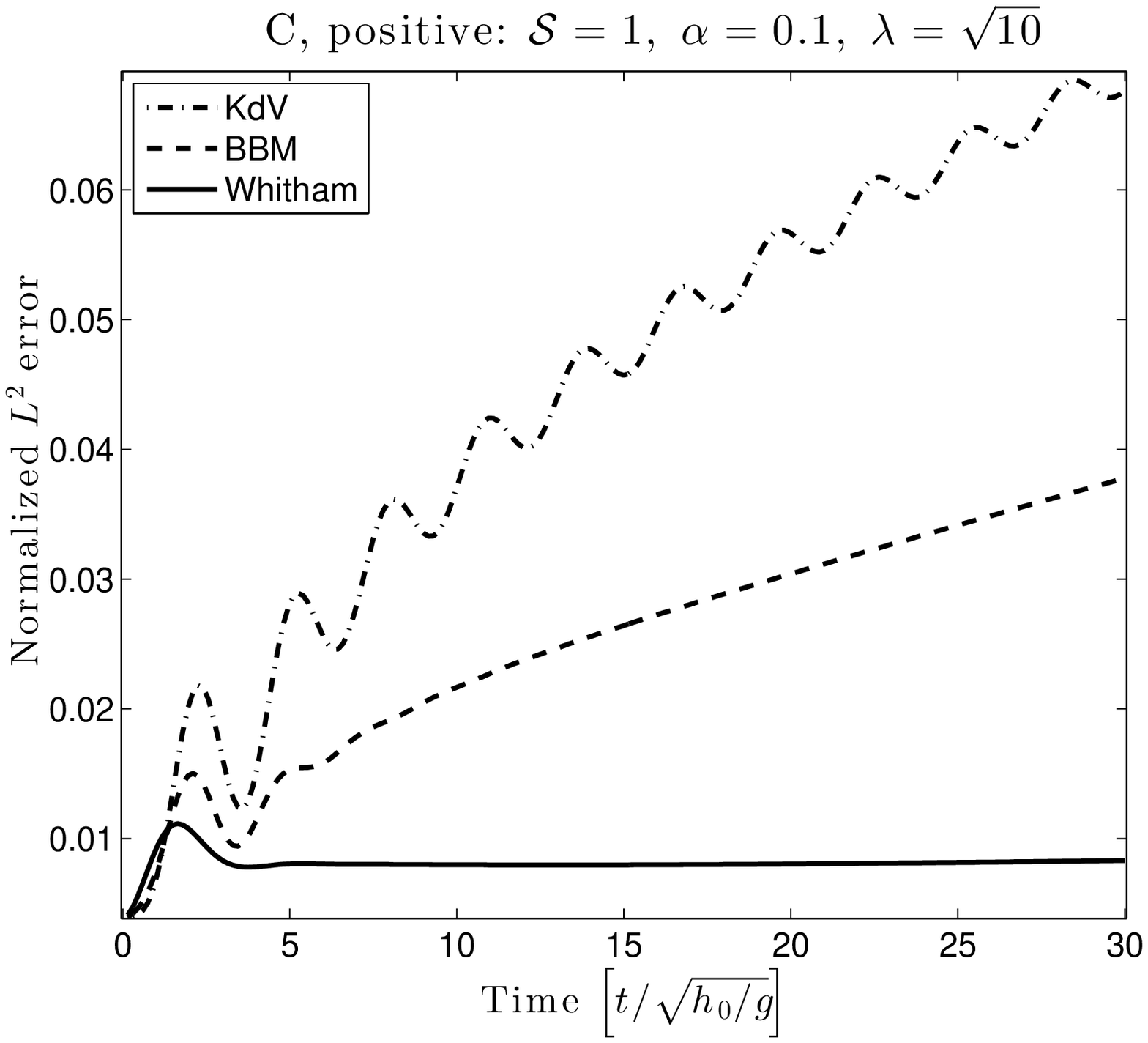}}~~~~
\subfigure{\includegraphics[width=0.39\textwidth]{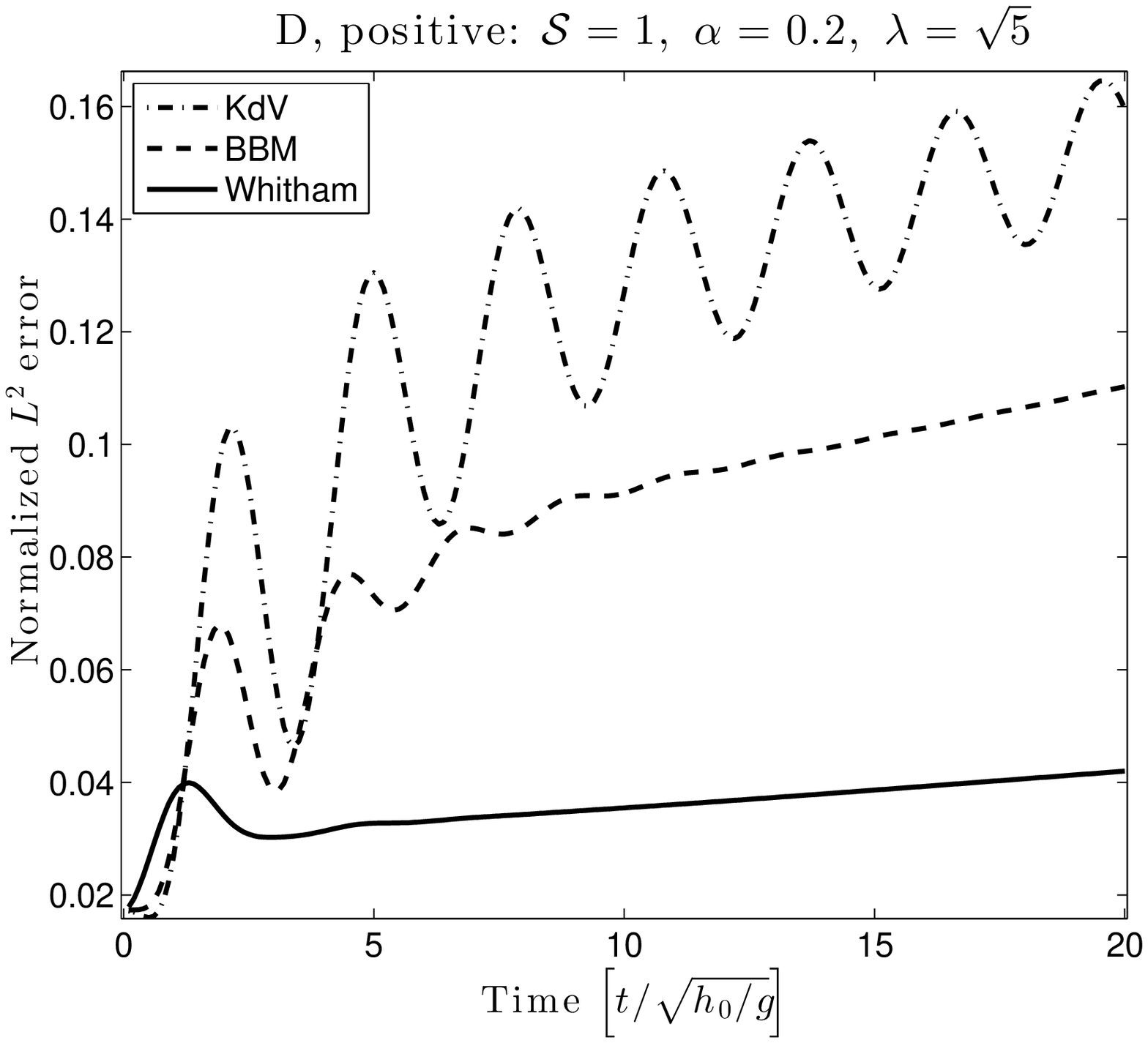}}
\subfigure{\includegraphics[width=0.39\textwidth]{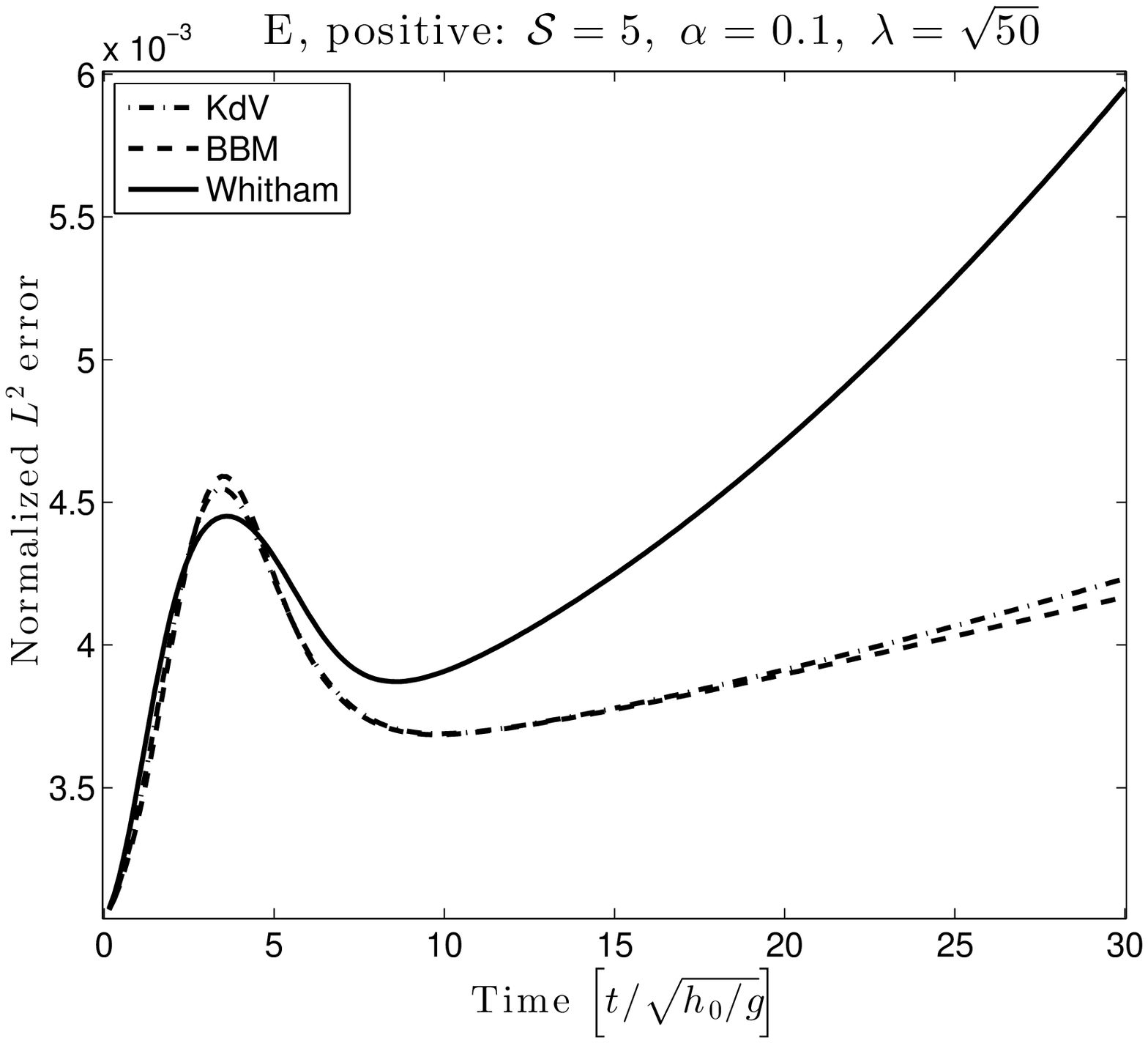}}~~~~
\subfigure{\includegraphics[width=0.39\textwidth]{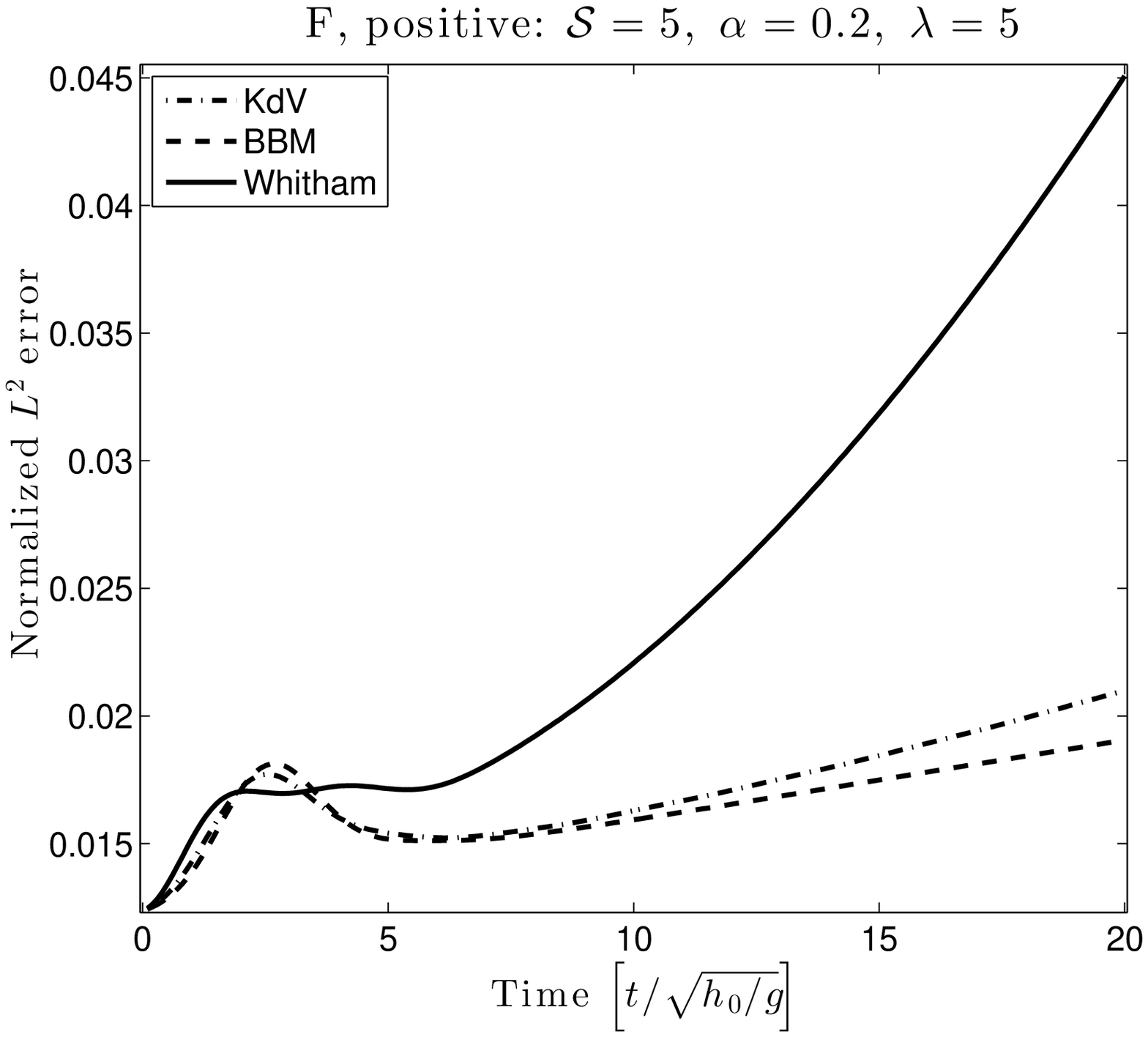}}
\caption{\small $L^2$ errors in approximation of solutions to full Euler equations 
by different model equations: cases A and B $(\mathcal{S} = 0.2)$,
cases C and D $(\mathcal{S} = 1)$, cases E and F $(\mathcal{S} = 5)$, positive.}
\label{fig3}
\end{figure}

\begin{figure}
\centering
\subfigure{\includegraphics[width=0.39\textwidth]{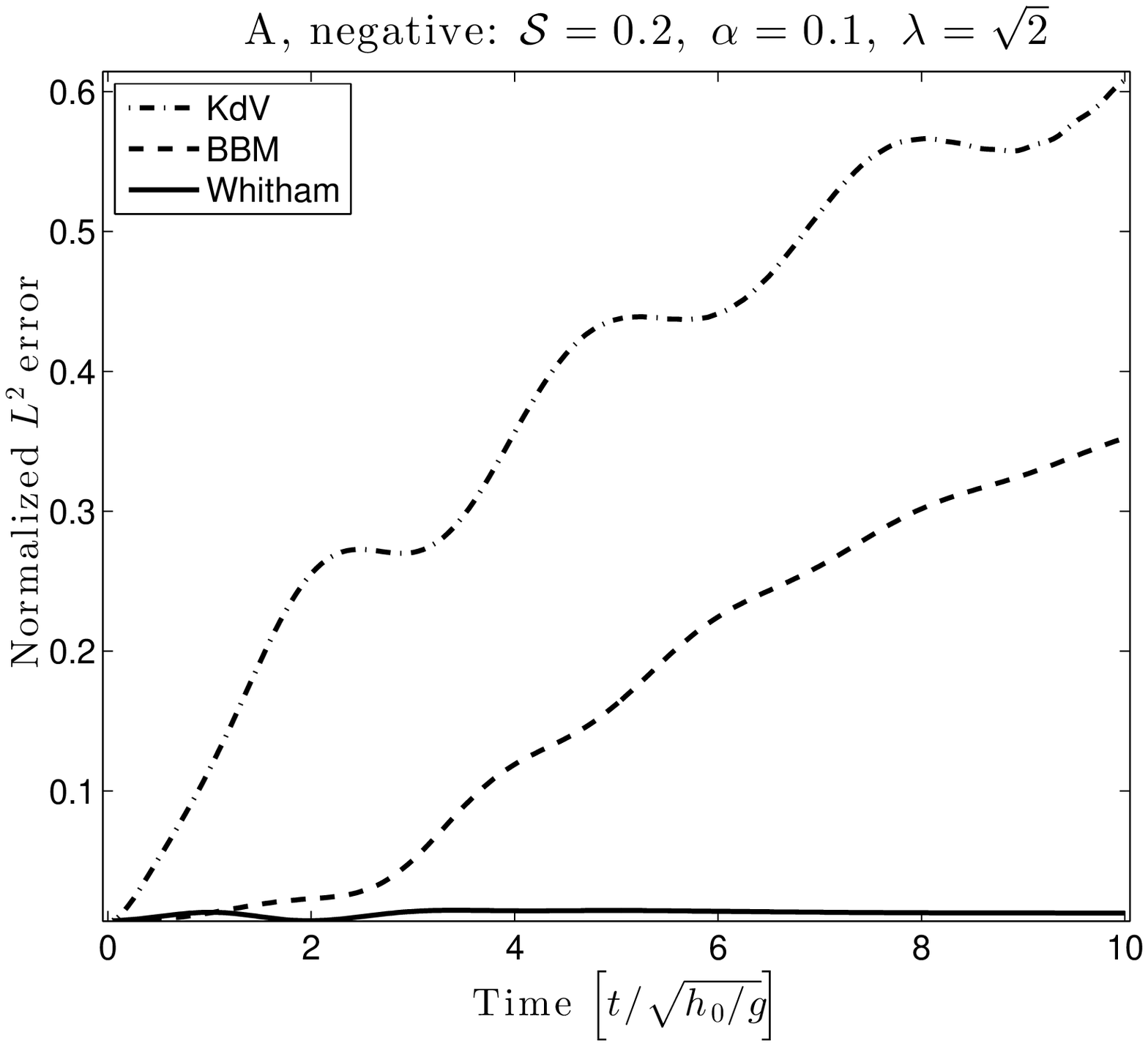}}~~~~
\subfigure{\includegraphics[width=0.39\textwidth]{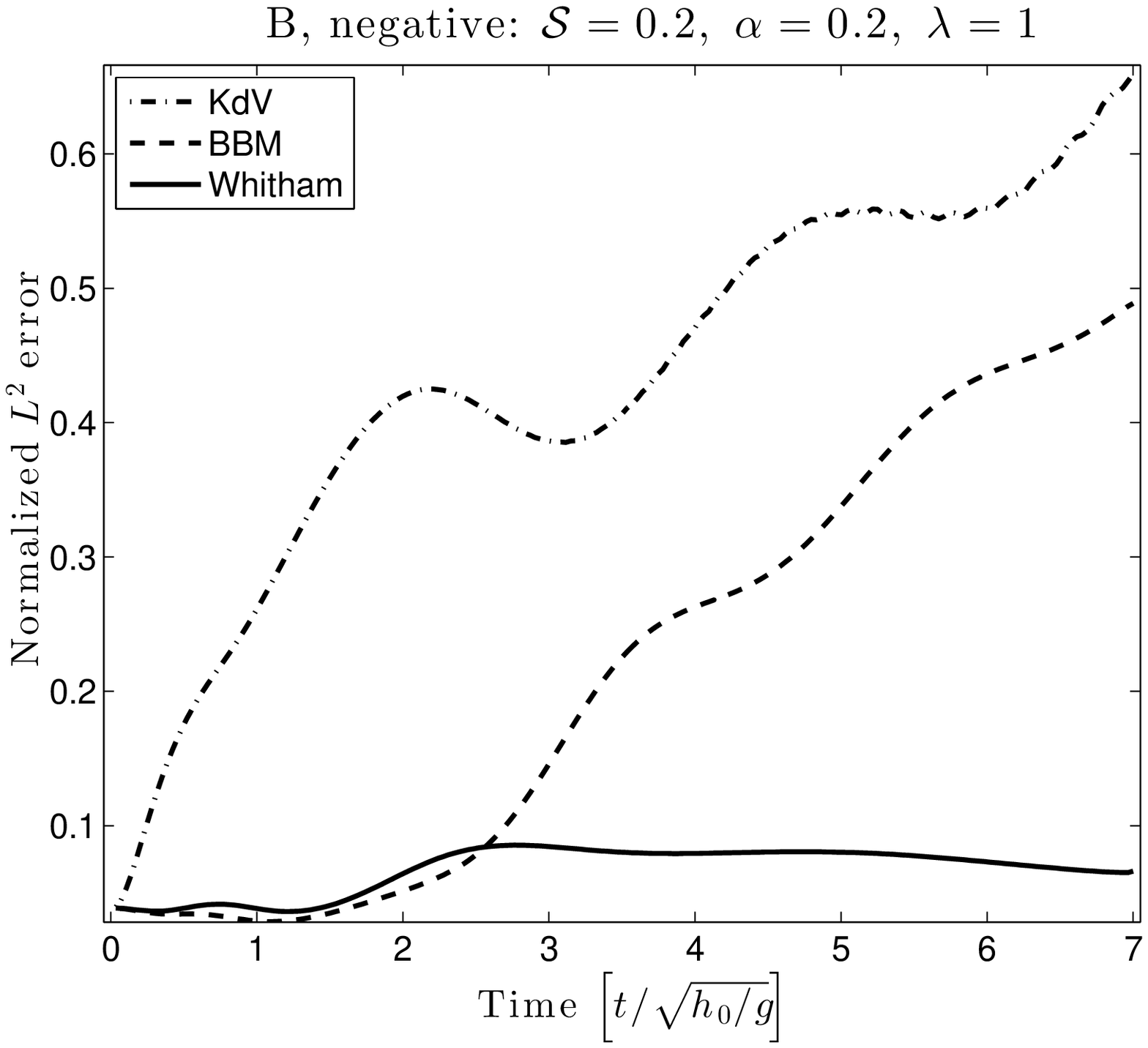}}
\subfigure{\includegraphics[width=0.39\textwidth]{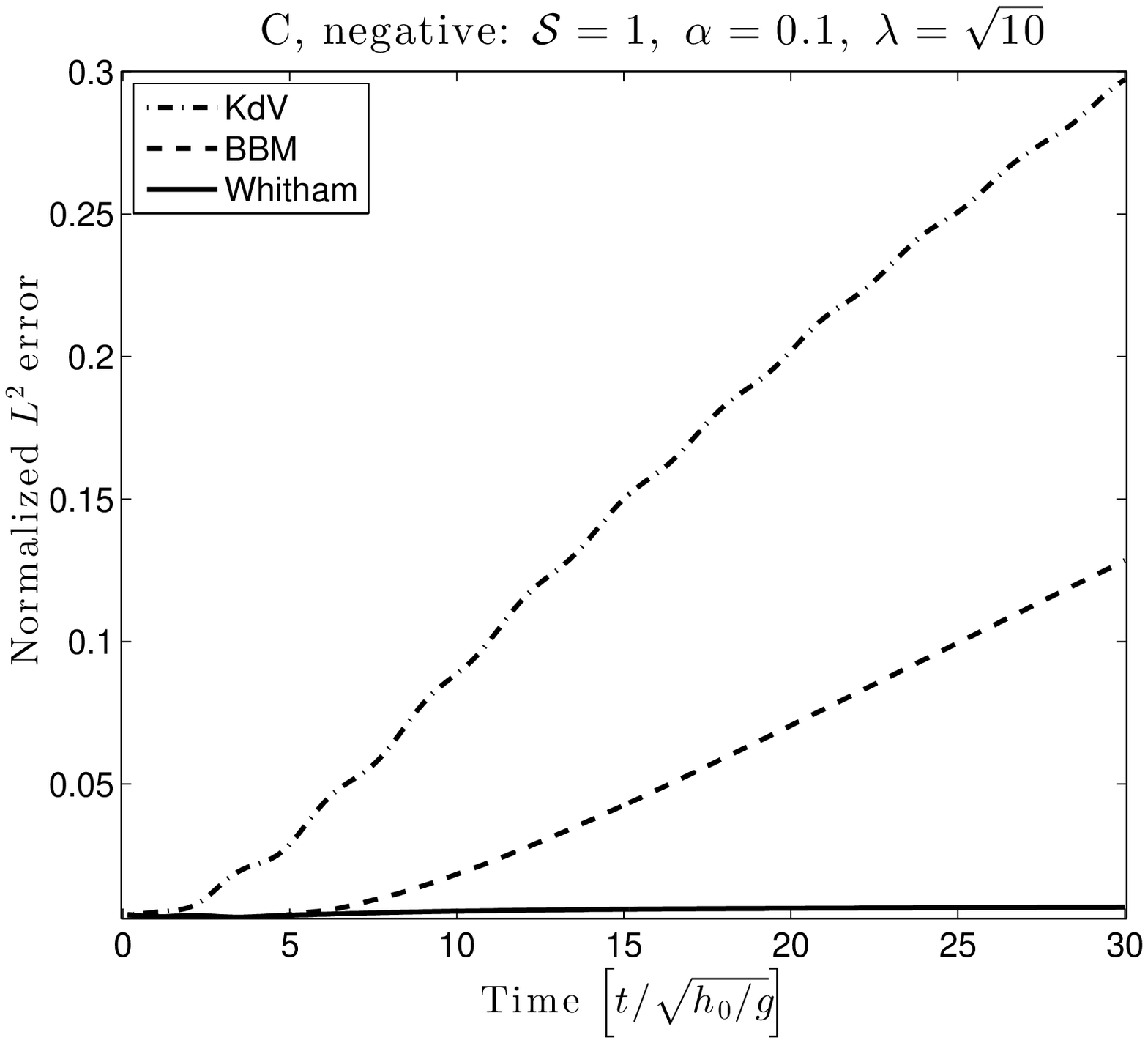}}~~~~
\subfigure{\includegraphics[width=0.39\textwidth]{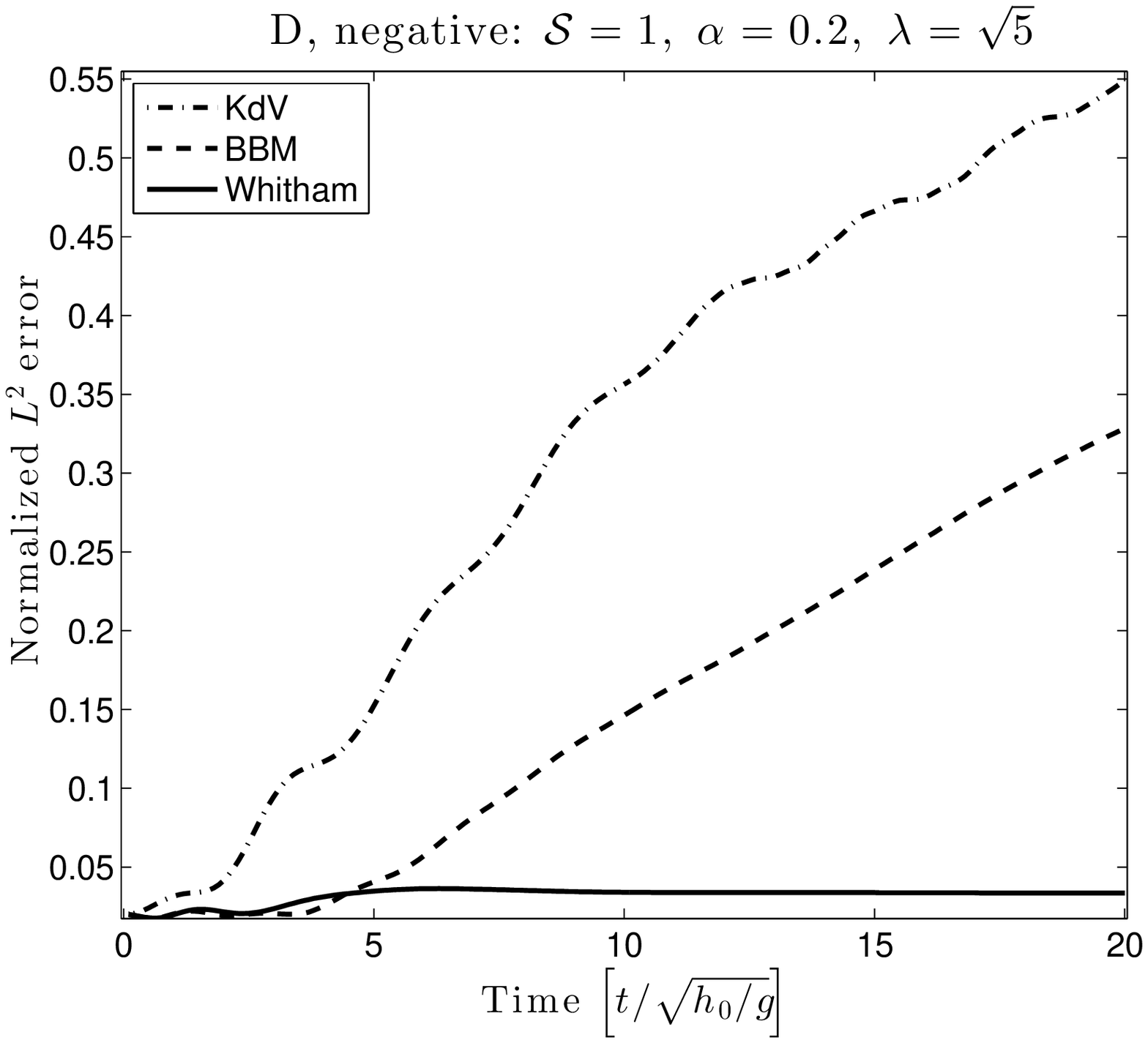}}
\subfigure{\includegraphics[width=0.39\textwidth]{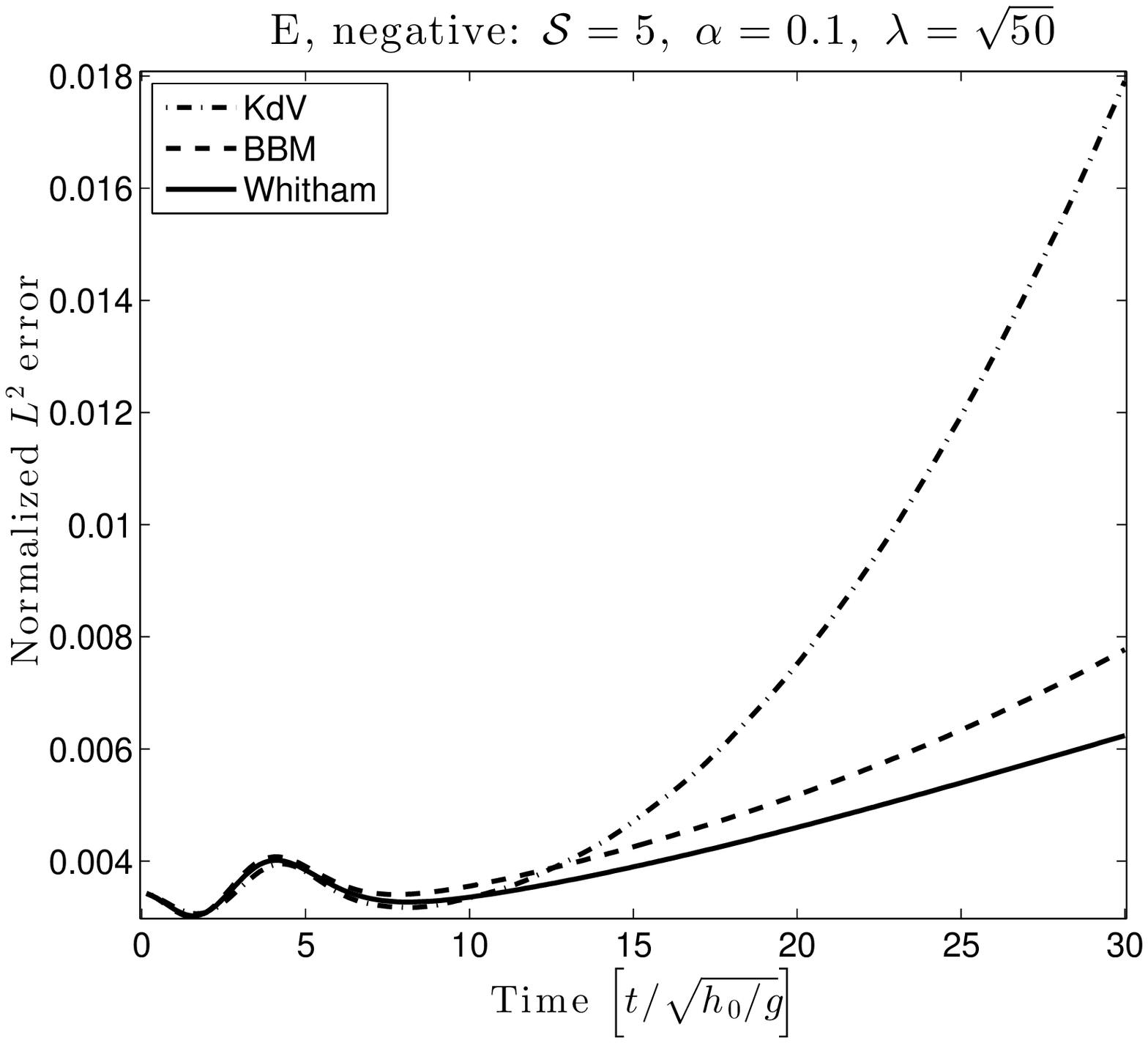}}~~~~
\subfigure{\includegraphics[width=0.39\textwidth]{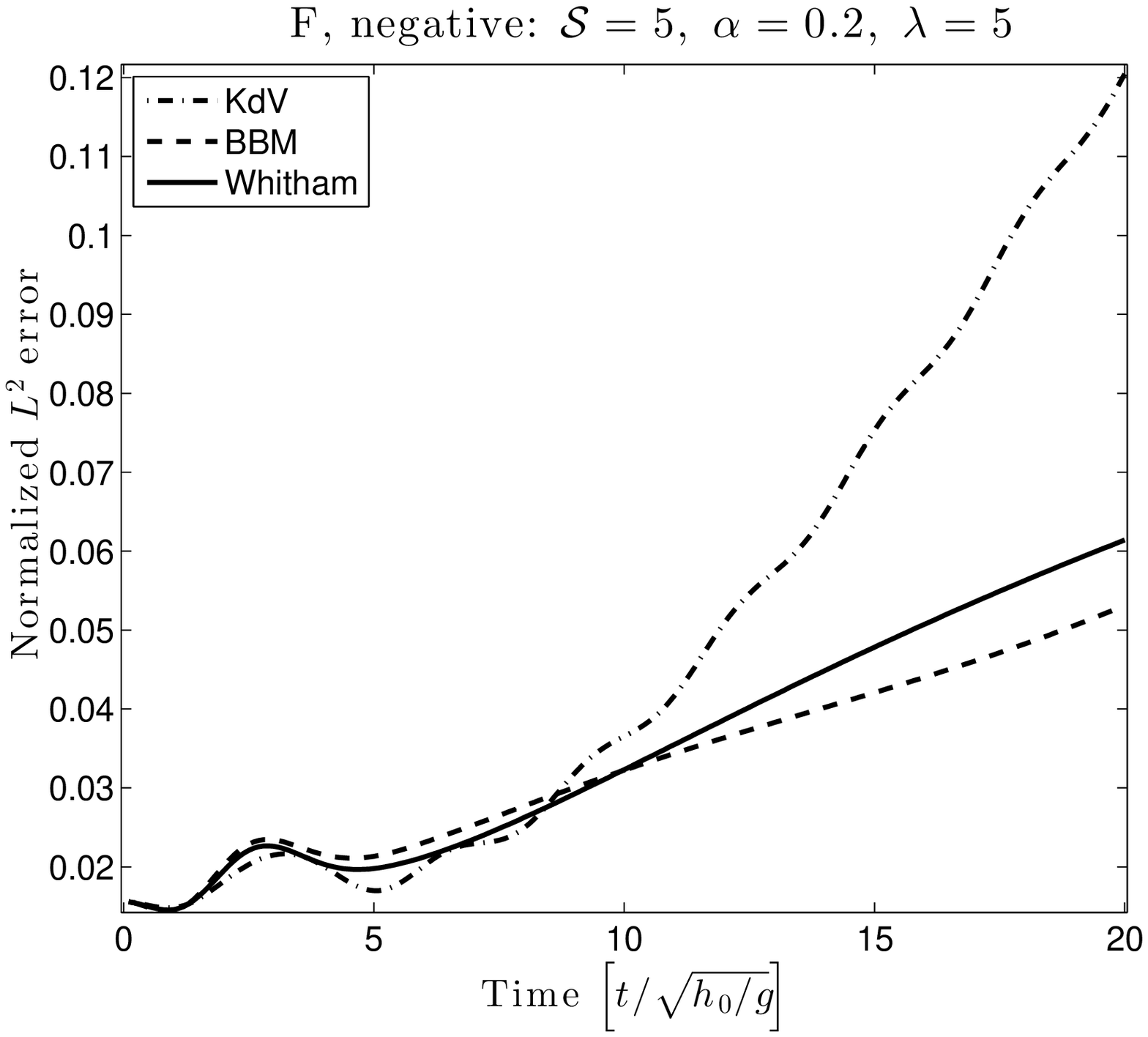}}
\caption{\small $L^2$ errors in approximation of solutions to full Euler equations 
by different model equations: cases A and B $(\mathcal{S} = 0.2)$,
cases C and D $(\mathcal{S} = 1)$, cases E and F $(\mathcal{S} = 5)$, negative.}
\label{fig4}
\end{figure}

In the center right panels of figures \ref{fig3} and \ref{fig4}, 
the computations from Figure \ref{fig2} are
summarized by plotting the normalized $L^2$-error between the KdV, BBM and Whitham,
respectively, and the Euler solutions as a function of non-dimensional time.
Using this quantitative measure of comparison, it appears that the Whitham
equation gives a better overall rendition of the free surface dynamics
predicted by the Euler equations.

In the center left panels of figures \ref{fig3} and \ref{fig4}, 
a similar computation with $\mathcal{S} = 1$, but smaller amplitude is analyzed. 
Also in these cases, it appears that the Whitham
equation gives a good approximation to the corresponding Euler solutions, and in
particular, a much better approximation than either the KdV or the BBM equation. 

Figures \ref{fig3} and \ref{fig4} show comparisons in several other cases of both positive
and negative initial amplitude, 
and Stokes numbers $\mathcal{S} = 0.2$, $\mathcal{S} = 1$ and $\mathcal{S} = 5$.
In most cases, the normalized $L^2$-error between the Whitham and Euler solutions is similar
or smaller than the errors between KdV, respective BBM and Euler solutions. 
The only case in this study in which the KdV and BBM equations outperform
the Whitham equation is in the case of very long waves (lower panels of Figure \ref{fig3}). 
In this case, we have $\mathcal{S} = 5$, and the main wave of the initial disturbance is positive.
However, even in this case, the Whitham equation yields approximations of the
Euler solutions which are similar or better than in the case of smaller wavelengths.
In addition, in the case of negative initial data, the performance
of the Whitham equation is on par with the KdV and BBM equations in the case when
$\mathcal{S} = 5$ (lower panels of Figure \ref{fig3}).

%
%
\section{Conclusion}
%
In this article, the Whitham equation \eqref{dimWhitham}
has been studied as an approximate model equation for wave motion
at the surface of a perfect fluid. Numerical integration of the
equation suggests that broad classes of initial data decompose
into individual solitary waves. The wavelength-amplitude ratio
of these approximate solitary waves has been studied, and it 
was found that this ratio can be described by an exponential
relation of the form
$\frac{a}{h_0} \sim e^{-\kappa(l/h_0)^\nu}$.
Using this scaling in the Hamiltonian formulation of the water-wave problem,
a system of evolution equations has been derived which contains the exact
dispersion relation of the water-wave problem in its linear part.
Restricting to one-way propagation, the Whitham equation emerged
as a model which combines the usual quadratic nonlinearity 
with one branch of the exact dispersion relation of the water-wave problem.
The performance of the Whitham equation in the approximation of
solutions of the Euler equations free-surface boundary conditions
was analyzed, and compared to the performance of the KdV and BBM
equations. It was found that the Whitham equation gives
a more faithful representation of the Euler solutions than either
of the other two model equations, except in the case of very long
waves with positive main part.

\vskip 0.05in
\noindent
{\bf Acknowledgments.}
This research was supported by the Research Council of Norway.

\end{document}